\documentclass[12pt]{iopart}

\usepackage{amssymb,bm,psfrag,epsfig}

\newcommand{\bms}{\bm{s}}
\newcommand{\bmsigma}{\bm{\sigma}}

\begin{document}
\title[A unified framework for Schelling's model of segregation]{A unified framework for \\Schelling's model of segregation}
\author{Tim Rogers, Alan J. McKane}
\address{Theoretical Physics Division, School of Physics \& Astronomy, \\The University of Manchester,
M13 9PL, UK}
\ead{tim.rogers@manchester.ac.uk, alan.mckane@manchester.ac.uk}
\begin{abstract}
Schelling's model of segregation is one of the first and most influential models in the field of social simulation. There are many variations of the model which have been proposed and simulated over the last forty years, though the present state of the literature on the subject is somewhat fragmented and lacking comprehensive analytical treatments. In this article a unified mathematical framework for Schelling's model and its many variants is developed. This methodology is useful in two regards: firstly, it provides a tool with which to understand the differences observed between models; secondly, phenomena which appear in several model variations may be understood in more depth through analytic studies of simpler versions.
\end{abstract}
\section{Introduction}
The Schelling model is one of the best known mathematical models in the social sciences. There are several reasons for this. Firstly, it is one of the oldest, having been proposed over forty years ago \cite{Schelling1969}. Secondly, it is one of the most easily described: in the model there are two types of individuals (agents) who tend to move if they find themselves in regions where the other type predominates. Thirdly, it proved very successful in illustrating a simple point, namely that only a slight homophilic bias is sufficient to cause wholesale segregation of the two types of agents \cite{Schelling1969}. Fourthly, this finding came at a time when the racial segregation that was occurring in US cities was at the centre of political discussions.\par
While these general aspects of the Schelling model are well known, the literature on the model since its introduction, and on its variants and generalisations, is surprisingly disjointed and unsystematic. Part of the reason for this is that an unusually large number of variations to the model have been proposed. Schelling himself produced several iterations of his first model \cite{Schelling1969,Schelling1971,Schelling1978}; a consequence of which is that there is no definitive ``Schelling model''. As computing power increased many researchers have simulated model variants related to their particular interests \cite{Zhang2004,Fossett2006,Stauffer2007,Fagiolo2007,Pancs2007,Benard2007,Barr2008,DallAsta2008,Singh2009,Gracia-Lazaro2009,OSullivan2009,Benenson2009,Fossett2009,Grauwin2009,Gauvin2009,Gauvin2010,Banos2010,Zhang2011}, finding segregation everywhere from the suburbs of Tel-Aviv \cite{Benenson2009} to the Sierpinski fractal \cite{Banos2010}.\par
A second reason for the lack of coherence of the literature has been that the model has attracted attention from researchers in several disciplines, who have tended to explore different aspects of the model, often choosing adaptations which are idiosyncratic to their own field of research. In particular, publications in the physics literature have gone to some considerable efforts to map the Schelling model onto systems already known to them, including liquids \cite{Vinkovic2006} and spin models \cite{Stauffer2007,Grauwin2009,Gauvin2009,Gauvin2010}. Whilst such analogies are interesting, they imply that the models ought to be studied from a within particular physical formalism, even though the corresponding assumptions, techniques and results may not necessarily be the most relevant to the interests of researchers in other fields.\par
Another striking feature of the literature on the Schelling model is the strong preference for numerical simulations over mathematical analysis. Aside from a few mathematical papers investigating limit states of deterministic versions of the model \cite{Pollicott2001,Gerhold2008}, analytical results on Schelling-like models are conspicuous in their absence. Surprisingly, this situation persists even in the physics literature where the typical models proposed are still too complicated to admit a successful theoretical treatment and must instead be simulated. Two recent exceptions to this are the works of Grauwin \textit{et al.}~\cite{Grauwin2009}, which maps a Schelling-like model onto a problem amenable to equilibrium statistical mechanics, and Dall'Asta \textit{et al.}~\cite{DallAsta2008}, with results including scaling laws for the development of clusters of agents.  \par
In this paper we initiate a comprehensive analysis of the Schelling model, its predictions and generalisations. We begin in Section 2 by constructing a unified mathematical framework for the broadest possible class of models of the Schelling type. We do this by using the ingredients of the models studied in the literature to date, by introducing elements motivated by models of other phenomena in the physical and biological sciences and by endeavouring to retain the essence of Schelling's original idea. The resulting construct encompasses the majority of the Schelling model variations proposed in the literature, enabling a systematic investigation of the relationship between the different models. \par
In simulations of Schelling model variations, discussed in Section 3, it is often found that the basic behaviour of segregation persists in a wide parameter range and is robust to the many adaptations made to the model. In light of this fact, the unified framework proposed here becomes a useful theoretical tool -- by exploring the full range of Schelling-class models, we are able to identify those which both exhibit the behaviour we are interested in and are well adapted to existing analytical tools. We demonstrate this methodology in Section 4, where a highly simplified Schelling-class model is used to develop an effective theory for the initial emergence of segregation, as measured by the density of unlike neighbours, providing a close fit to data from simulations from Schelling model variants.  In a forthcoming paper, similar techniques are applied to give an analytical treatment of the more complex phenomena of jamming and pattern formation exhibited by certain Schelling-class models \cite{Rogers2011iii}.
\section{General model structure}
\subsection{Model components}
As discussed in the introduction, there is surprising variety in the models which have been simulated and studied under the heading of Schelling's model. It is natural, then, to ask precisely what constitutes a Schelling model, and how changes in the model specification may or may not affect the behaviour observed. To facilitate this discussion, we will develop in this section a general mathematical framework which encompasses most (if not all) of the models studied in the literature.\par
Certain broad features of the model are common to almost all approaches: agents of two types, which we label $A$ and $B$, are allowed to move in some space, without being created or destroyed; agents are motivated to move on the basis of their level of satisfaction, which in turn is decided by the makeup of their neighbourhood. \par
We propose that the formulation of a model with this overarching structure can be reduced to the choice of four key components: the space in which agents move, their initial numbers and arrangement, the function which decides their satisfaction, and lastly the mechanism for selecting if and when a particular move should take place. We discuss these components in turn.
\subsubsection{Network}
The concept of segregation is inherently spatial; to recognise two groups as separate we must have some notion of when agents are close to each other and when they are not. In Schelling's original work this distance is geographical, with the city divided into a grid of residences and closeness defined by one's neighbours in the grid \cite{Schelling1969}. One could equally well consider other ideas of distance \cite{Fagiolo2007,Banos2010}, for example in terms of interpersonal relationships. \par
Mathematically, this situation is most easily formalised in terms of a network. We consider a collection of $N$ sites, joined by some number of edges. We label the sites by the numbers $1,\ldots,N$. If two sites $i$ and $j$ are joined by an edge, we say that $i$ and $j$ are neighbours in the network. The neighbourhood of $i$ is defined to be the set of all neighbours of $i$ and is written $\partial i$. For simplicity we consider networks without multiple edges between the same sites and without edges joining a site to itself. \par
At any given moment, a site may be occupied by at most one agent, or it may be vacant. We encode the status of site $i$ in an integer variable $\sigma_i$ by setting 
\begin{equation}
\sigma_i=\cases{1\quad&\textrm{if site $i$ is occupied by an agent of type $A$}\\-1&\textrm{if site $i$ is occupied by an agent of type $B$}\\0&\textrm{if site $i$ is vacant.}}
\label{statevector}
\end{equation}
Using these variables, the state of the whole system at a given time (that is, the location of each of the agents in the network) is specified by the vector $\bmsigma=(\sigma_1,\ldots,\sigma_N)$. \par
The choice of encoding (\ref{statevector}) is a common approach in the physics literature (see, for example, \cite{DallAsta2008,Gauvin2009,Gauvin2010}) and is useful mainly for mathematical reasons, as it provides simple formulae for several quantities we may be interested in. For example, for any two sites $i$ and $j$,
\begin{equation*}
\sigma_i\sigma_j=\cases{1\quad&\textrm{if $i$ and $j$ are occupied by agents of the same type}\\-1&\textrm{if $i$ and $j$ are occupied by agents of different types}\\0&\textrm{if either site is vacant.}}
\end{equation*}
Note that we have made the sites, rather than the agents, into the primary objects of interest -- all agents of the same type are seen to be equivalent, and we are only concerned with which sites they occupy. \par
In some of the more complicated Schelling-inspired models agents are endowed with additional properties beyond their race, for example wealth and social status \cite{Benard2007}. This is also true of the residences, which may, for example, have a price \cite{Yin2009}. These additional factors can be incorporated into the present framework by taking vector-valued site status variables $\sigma_i$, in which case each property (either of the site or of the occupying resident) is specified by an entry of the vector. 
\subsubsection{Initial condition}
In developing his models, Schelling was concerned with the mechanisms driving the spontaneous segregation of an initially well-integrated society. It follows that the initial state of the model should reflect a society which is not segregated -- Schelling himself chose to place agents at random without any bias \cite{Schelling1969}.\par
Since agents are neither created nor destroyed during the course of the model dynamics, the initial condition fixes permanently the number of agents of each type, which may or may not be equal. Also specified is the number and location of the vacancies which in most Schelling-class models facilitate the movement of the agents \footnote{Almost every model cited here follows this rule. An interesting exception is the model Zhang \cite{Zhang2004} in which agents directly exchange location. The same reference is also unusual in employing a non-random initial condition.} \par
In many versions of the model, the fraction of vacant sites plays an important role in the dynamics; this is particularly the case if it is taken to be small, as discussed in \cite{Gauvin2009} and \cite{Edmonds2005} for example. We denote this quantity by $\rho$, which may be written as a function of the state vector $\bm{\sigma}$,
\begin{equation*}
\rho=\frac{1}{N}\sum_i\big(1-|\sigma_i|\big)\,.
\end{equation*}
In the language of the framework for Schelling-class models we are developing here, choosing an initial condition amounts to specifying the state $\bmsigma$ of the system at time $t=0$. Formally, to employ a random initial condition, we should take the starting state to be a random variable with some specified law $\mu:\{-1,0,1\}^N\to[0,1]$. In practice, however, it is more convenient to specify the choice of initial condition with words, as an explicit formula for $\mu$ will rarely provide any great insight.\par
To summarise, the initial condition is specified by the state vector $\bmsigma$ at time $t=0$, which fixes permanently the number of agents of each type, as well as the fraction of vacant sites, denoted by $\rho$. In most models the initial condition will be chosen at random, usually without any bias in the placement of agents of different types. 
\subsubsection{Satisfaction function}
In all versions of Schelling's model, the movement of the agents is motivated by a measure of how satisfied an agent is with its current location and/or how satisfied it would be with a potential future location. The spirit of Schelling's work is captured by the general heuristic that an agent's satisfaction should be low if its neighbours are predominantly of the opposite type, though many different interpretations of this requirement have been used in the past. \par
Introduce the vector $\bms=(s_1,\dots,s_N)$, where the entry $s_i$ is a real number between $0$ and $1$, encoding the satisfaction of the agent occupying site $i$ (if $i$ is vacant, we set $s_i=0$). \par
Given the variety of different satisfaction functions used in the literature, our general framework should be broad enough to include any sensible choice. We make no restrictions other than to specify that satisfaction should depend only on the number of like and unlike neighbours an agent has. Mathematically, this means that $s_i$ is a function of the numbers $\sigma_i\sigma_j$, for $j\in\partial i$. \par
Frequently in the literature the satisfaction $s_i$ of an occupied site $i$ is taken to depend on the fraction of occupied neighbours of that site containing agents of the opposite type. We denote this quantity by $x_i$, where $0\leq x_i \leq 1$, and we write $x_i=0$ if site $i$ is either vacant itself or surrounded by vacant sites, otherwise
\begin{equation}
x_i=\frac{\sum_{j\in\partial i}\big(|\sigma_i\sigma_j|-\sigma_i\sigma_j\big)}{2\sum_{j\in\partial i}|\sigma_i\sigma_j|}\,.
\label{x_i}
\end{equation}
\subsubsection{Transfer probabilities}
Agents move either by finding a vacant site to relocate to (leaving their starting site vacant), or in some models by directly swapping with another agent. In either case, a move will result in two entries of the state vector being exchanged. To express this mathematically, we introduce the notation $\bmsigma^{(ij)}$ for the state vector which would result from $\bmsigma$ if the contents of sites $i$ and $j$ were swapped. We also write $\bms^{(ij)}$ for the satisfaction levels after the change. Note that this is not the same as swapping the entries of $\bms$ in positions $i$ and $j$: in general $s_i^{(ij)}\neq s_j$, as $s_i^{(ij)}$ specifies how satisfied the agent currently in site $i$ would be if it were to move to site $j$.\par
Although we have decided that agents are motivated to move by their level of satisfaction, we have not established how a swaps should be selected or how likely a certain swap is to take place. In short, we must choose the transfer probabilities $T_{ij}(\bmsigma)$, giving the likelihood that sites $i$ and $j$ will be selected (in that order) and that their contents will be swapped. \par
For the model to be well defined, the transfer probabilities must all be non-negative and together satisfy for all $\bmsigma$,
\begin{equation*}
\sum_{i,j}T_{ij}(\bmsigma)=1\,,
\end{equation*}
Write $P(\bmsigma,t)$ for the probability that the system is in state $\bmsigma$ at time $t$. The initial condition specifies $P(\bmsigma,0)$, and for $t>0$, the evolution of the system is determined by the transfer probabilities according to
\begin{equation*}
P(\bmsigma,t+1)=\sum_{i,j}T_{ij}(\bmsigma^{(ij)})\,P(\bmsigma^{(ij)},t)\,.
\end{equation*}
Whilst in theory any variant of the Schelling model can be described in this way, it is common in the literature (from \cite{Schelling1969} onwards) to define the model dynamics in terms of an algorithm, rather than an explicit transfer probability. Unfortunately, the great number of different algorithms suggested makes it difficult to place meaningful limitations on the structure of $T$ without ruling out potentially interesting models. This task is necessary, however, as without constraints on $T$ almost any dynamics could occur and it is not at all clear how the nature of the satisfaction function and network structure is to influence the behaviour of the model. This is one of the central difficulties in formulating a useful mathematical framework for the study of the many variants of the Schelling model. \par
Our solution is to specify that $T_{ij}(\bmsigma)$ should be taken as the product of three components
\begin{enumerate}
\item The probability of selecting the agent in site $i$ to be given the opportunity to move \\(zero if $i$ is vacant, and non-increasing in $s_i$)
\item The probability of selecting site $j$ as the destination of the move \\(non-increasing in the distance from $i$ to $j$) 
\item A measure of how desirable site $j$ is to the agent at $i$ \\(non-decreasing in $s_j^{(ij)})$
\end{enumerate}
It is hoped that this structure is simple and restrictive enough to make clear the way in which the agents act selfishly to pursue their own satisfaction, whilst remaining broad enough to include a great many of the different model variants.
\subsection{Examples}
In the previous subsection we developed a unified mathematical framework for Schelling-class models, based on the specification of four components: a network, an initial condition, a satisfaction function, and a formula to decide the transfer probabilities. We now give some examples of Schelling-class models studied previously, showing how the various model specifications fit within our framework.
\subsubsection{Schelling}
As described in the introduction, the residences in Schelling's original two-dimensional model \cite{Schelling1969} of a city were arranged in a grid. In our framework this setting corresponds to a two-dimensional lattice in which each site has eight neighbours (i.e.~one in each horizontal, vertical and diagonal direction). The initial condition specified by Schelling has an equal number of agents of each type placed randomly on the network, leaving a proportion $\rho$ of the sites vacant. Schelling's choice for the satisfaction function for occupied sites is given simply by
\begin{equation*}
s_i=\cases{1&if at least half of the neighbours of $i$ are of the same type,\\0&otherwise.}
\end{equation*} 
The transfer probabilities are all zero except for single pair $i,j$ with $T_{ij}(\bmsigma)=1$, where $i$ is the next unsatisfied site to be updated and $j$ is the nearest vacant site which would satisfy the agent in site $i$. Schelling was not entirely explicit about the precise definitions of `next' and `nearest' to be used.
\subsubsection{Pancs and Vriend}
In \cite{Pancs2007}, Pancs and Vriend use the same network and initial condition as Schelling, but make key changes to the satisfaction function and transfer probabilities. The main alteration made is the assumption that agents select their destination by assessing each vacant site to find the one which maximises a utility function -- this kind of `best response' dynamics is a common modelling paradigm in the economics literature.  \par
In the model of Pancs and Vriend, the satisfaction of an occupied site $i$ is determined as a function $u$ of the fraction of occupied neighbouring sites which contain an agent of the opposite type (recall that we denote this quantity by $x_i$). The number $u(x_j^{(ij)})$ represents the utility of the site $j$ to the agent currently in site $i$. The dynamics of the model are described by the following rule: at each timestep an occupied site $i$ is chosen at random, the agent there moves to a vacant site $j$ which is chosen at random from those maximising $u(x_j^{(ij)})$. The transfer probabilities resulting from this procedure may be written explicitly using the slightly complicated expression
\begin{equation*}
T_{ij}(\bmsigma)=\left(\frac{|\sigma_i|}{1-\rho}\right)\frac{(1-|\sigma_j|)\displaystyle\mathbb{I}\left\{u(x_j^{(ij)})=\max_k \,u(x_k^{(ik)})\right\}}{\displaystyle \bigg.\sum_l(1-|\sigma_l|)\mathbb{I}\left\{u(x_l^{(il)})=\max_k \,u(x_k^{(ik)})\right\}}\,.
\end{equation*}
Here we have used the indicator function $\mathbb{I}$, whose output is one if the argument is a true statement and zero if it is false.
\subsubsection{Gauvin \textit{\textit{et al.}}}
The authors of \cite{Gauvin2009} make much simpler model definitions, with the aim of suggesting a link between the Schelling model and the Blume-Emery-Griffiths spin model \cite{Blume1971}. Once again the same network and initial condition is used, with a range of values of the density of vacancies $\rho$. A second parameter, here called $\tau$, is introduced, giving the maximum fraction of unlike neighbours that an agent will tolerate. The satisfaction function and transition probabilities are then given by
\begin{equation*}
s_i=\cases{1&if $\quad x_i<\tau$\\0&otherwise,}
\end{equation*}
and
\begin{equation*}
T_{ij}(\bmsigma)=\frac{|\sigma_i|}{N(1-\rho)}\,\frac{1-|\sigma_j|}{N\rho}\,s^{(ij)}_j\,.
\end{equation*}
\subsubsection{Laurie and Jaggi}
For a final example we turn to the sociology literature and the model of Laurie and Jaggi \cite{Laurie2003}, which was introduced to study the effect of increasing the range of `vision' of the agents. This factor is incorporated into the model by expanding the size of neighbourhood of a residence from the eight surrounding sites to include all sites within a distance of $R$; by increasing this parameter the agents have a wider view of the sites near them in the grid. \par
The initial condition is randomised as usual, but with the possibility of an uneven split between the numbers of $A$ and $B$ agents, controlled by a parameter $c$ giving the fractional size of the minority. The satisfaction function used is simply the fraction of unlike agents in the neighbourhood. Some slight changes are also made to the dynamical rules. At each timestep an agent is chosen; if its satisfaction is below a threshold $p$, it will make up to $N\rho$ attempts to move to a randomly selected vacancy -- a given attempted move is completed if it results in an increase to the agent's satisfaction. The transfer probabilities for this scheme are found by summing over number of unsuccessful attempts:
\begin{equation*}
\hspace{-71pt}T_{ij}(\bmsigma)=\frac{|\sigma_i|}{N(1-\rho)}\,\frac{1-|\sigma_j|}{N\rho}\Theta\Big[\,p-s_i\Big]\Theta\Big[s_j^{(ij)}-s_i\Big]\sum_{n=1}^{N\rho}\left(1-\frac{1}{N\rho}\sum_k\Theta\Big[s_k^{(ik)}-s_i\Big]\right)^{(n-1)}\hspace{-4pt},
\end{equation*}
where the $\Theta$ function gives 1 if its input is positive and 0 otherwise. 

\section{Simulation of existing models}
\subsection{Measuring segregation}
Before reporting the results of simulations of whichever version of the Schelling model one is interested in, it is first necessary to consider how the data are to be distilled into a format which provides useful quantitative information. Schelling himself, and many others since, have chosen to present diagrams (or latterly screenshots) showing the arrangement of agents and vacancies at certain times. These images typically show the emergence of patterns of agents of the same type forming domains of various shapes and sizes. The patterns observed are striking and are no doubt responsible for generating much interest in the model, however they are not enough on their own to provide quantitative information about the dynamics of the system, or to compare different parameter or model choices. In particular, this method of presenting data becomes very much less useful if the underlying network is anything other than a square lattice, as in \cite{Fagiolo2007,Banos2010}.\par
More sophisticated analyses of the behaviour of the system may be undertaken by considering appropriate numerical statistics which capture some feature of the state vector. The statistics usually considered fall broadly into two categories: those which count the frequency of certain local configurations of agents, and those which observe the global state of the system. We discuss the options in turn.\par
There are numerous ways to summarise the makeup of the neighbourhood of an individual agent: the ratio of like to unlike, fraction of non-vacant neighbouring sites occupied by like or unlike agents, the difference between the number of like and unlike agents, and so on. If such a local measurement is taken at each site and the result averaged over the whole network, one obtains an aggregate measure for the level of segregation in the system. We choose here to focus on interface density, a frequently studied quantity (for example in \cite{DallAsta2008,Gauvin2010,Banos2010}) which is a good representative of local statistics of this type. We define the interface density to be
\begin{equation}
x=\frac{\displaystyle\textrm{number of edges between agents of opposite types}}{\displaystyle\textrm{number of edges between agents of any type}}\,.
\label{def_x}
\end{equation}
Note that $x$ is the average over the whole network of the local quantity $x_i$ introduced in equation (\ref{x_i}).\par
On a global scale, many Schelling-class models develop large regions filled with agents of a single type. This behaviour is known as clustering; it might be expected that the distribution of cluster sizes and their shape could be used to characterise model variants. Many previous studies have numerically investigated the emergence of clusters (examples include \cite{Pancs2007,Gauvin2009,Singh2009} and many others), however it is unfortunately very difficult to make analytical progress in understanding cluster size distributions, even in relatively simple spin models in statistical physics \cite{Bray1994}. For this reason we do not focus on cluster properties in the present study.
\subsection{Simulation results}
As we have seen, many different versions of the Schelling model fit within the mathematical framework defined in the previous section, each of which can reasonably claim to describe a simplified mechanism for the emergence of segregation. Since these models all seek to describe the same phenomenon, one would hope that they do not exhibit wildly different behaviour (at least away from the extremes of their parameter space). We check this now by simulating several different Schelling-class models and comparing the time evolution of the interface density in each model.\par
\begin{figure}
\includegraphics[width=0.9\textwidth, trim=0 0 0 0]{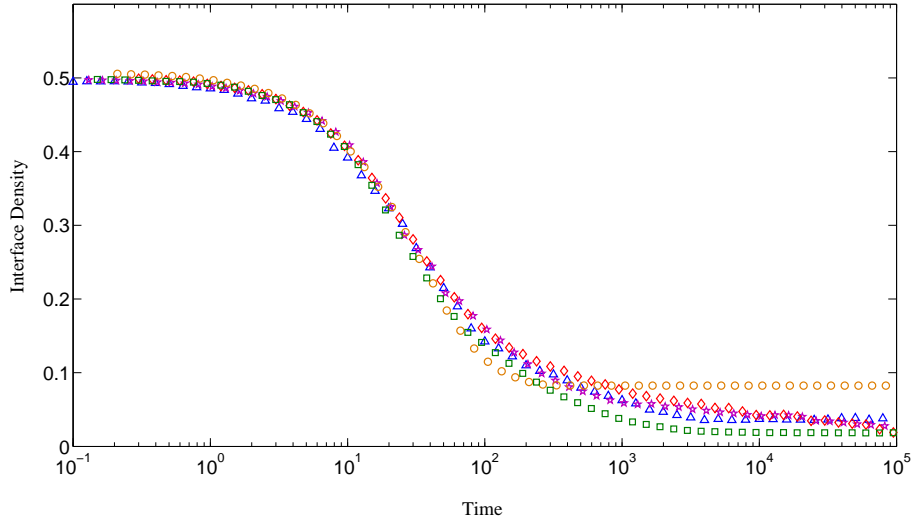}
\caption{Time evolution of the interface density in simulations of various Schelling-class models. Orange circles -- Laurie and Jaggi \cite{Laurie2003}, red diamonds -- Fagiolo \textit{et al.} \cite{Fagiolo2007}, purple stars -- Schelling \cite{Schelling1969}, blue triangles -- Gauvin \textit{et al.} \cite{Gauvin2009}, green squares -- Pancs and Vriend \cite{Pancs2007}. See the main text for model details and parameter values.}
\label{existing_models}
\end{figure}

The models have been chosen to provide some variation in three of the four model components: the network, the satisfaction function, and the transfer probabilities. They are
\begin{enumerate}
\item Schelling's original 2D model \cite{Schelling1969} on a toroidal grid of $N=10,000$ sites with vacancy density $\rho=0.1$
\item The best response model of Pancs and Vriend \cite{Pancs2007} on a toroidal grid of $N=10,000$ sites with vacancy density $\rho=0.1$ and utility function
\begin{equation}
u(x)=\cases{0.1+1.8\,x&if$\quad x\leq1/2$\\0&otherwise}
\label{PV03}
\end{equation}
This choice is a particularly interesting one as it implies that agents would be most satisfied in a mixed environment, however, in simulations segregation still emerges.
\item The model of Gauvin \textit{et al.} \cite{Gauvin2009} on a toroidal grid of $N=10,000$ sites with vacancy density $\rho=0.1$ and tolerance $\tau=0.6$
\item The model of Laurie and Jaggi \cite{Laurie2003} on a toroidal grid of $N=10,000$ sites with neighbourhood radius $R=2$, vacancy density $\rho=0.1$, an equal number of each type of agent ($c=1/2$), and threshold parameter $p=0.9$.
\item A model of Fagiolo \textit{et al.} \cite{Fagiolo2007} on a small-world network of $N=10,000$ sites and $4N$ edges with rewiring probability $0.2$ (see \cite{Fagiolo2007} for details), vacancy density $\rho=0.1$, and a Schelling-type satisfaction function.
\end{enumerate}
A plot of the time evolution of the interface density for a single simulation run of each model is shown in Figure \ref{existing_models}. In each case time has been rescaled by a factor of $N^{-1}$ to account for system size, and again by an amount to align the curves for comparison.\par
There are two regimes visible in this figure. In short to medium timescales, the simulation results show the rapid emergence of segregation in the models which, despite significant differences in their specification, follows a single characteristic curve. This common behaviour persists until the interface density has dropped below 0.1, by which stage the system is already in a quite strongly segregated state. Whilst the models considered may exhibit more unusual behaviour in certain extremes of their parameter spaces, the results shown here are typical for a fairly broad range of `reasonable' parameter choices, suggesting that the shape of curve seen above is quite robust to changes in model specification and parameters.  Consequently, it can be argued that if one is interested in studying this behaviour then there is considerable scope to vary the model definitions whilst keeping the analysis relevant.\par
After this initial segregation forming period, the different models begin to disagree at large times as each relaxes to an equilibrium state which depends upon the model specification, parameter choice and system size. There is variation in the nature of the equilibrium and the mechanism of relaxation. Some models (such as that of Laurie and Jaggi \cite{Laurie2003}) arrive at a stable configuration which does not change, either because every agent is satisfied or because no acceptable move exists; for certain models these limit states have been studied mathematically\cite{Pollicott2001,Gerhold2008}. Other models, particularly those which allow random moves that lower the satisfaction of the agent involved (for example Gauvin \textit{et al.} \cite{Gauvin2009}) reach something resembling a thermal equilibrium composed of many similar states. Some theoretical insight may be gained into the long-time processes at work in this case by considering the dynamics of moving groups of agents, which slowly form larger and larger clusters \cite{DallAsta2008,Gauvin2009}.
\section{Analytical treatment}
Most of the work on the Schelling model and its variants, even in the physics literature, has been based mainly on numerical analyses of simulations. One possible reason for the relative scarcity of analytical results is the complexity of the models usually considered and hence the apparent difficulty of undertaking a theoretical analysis. To make progress in a situation like this the traditional theoretical physics approach is to choose a particular behaviour or phenomenon to investigate, and seek out new versions of the model which capture the essential character of the problem, yet are simple enough to study analytically. \par
In this section we demonstrate this principle by analysing an extremely simple Schelling-class model, allowing us to write an effective theory for the emergence of segregation observed in the more complex models simulated in the previous section. 
\subsection{Construction of a simple model}
With the aim of analytically studying the emergence of segregation, we seek to construct the simplest Schelling-class model we can. As shown in the previous section (and also in \cite{Fagiolo2007}) variations in the structure of the underlying network do not appear to greatly alter the behaviour of the model in the main parameter regime. Moreover, it is frequently observed that network effects can greatly complicate the analysis of stochastic systems \cite{Rogers2011i}. With this in mind, we suggest to dispose of the network almost entirely, choosing instead to group sites in pairs so that each has exactly one neighbour. This can be thought of as an abstraction of the network in which the neighbourhood of a site is replaced with a single representative neighbour. \par
With only one neighbour per site, specifying a satisfaction function amounts to picking numbers $u,v\in[0,1]$ and setting
\begin{equation*}
s_i=\cases{u&if $i$'s neighbour is of the same type\\v&otherwise.}
\end{equation*}
For our model to carry the ethos of Schelling's, we set $u>v$.\par
Continuing the pursuit of a very simple model, we take an initial condition with equal numbers of agents of each type, placed randomly, with \textit{no vacancies}. In each time step, a randomly selected agent is allowed to move by swapping places with another (again selected at random), according to its satisfaction before and after the swap. Specifically, we set
\begin{equation}
T_{ij}(\bmsigma)=\frac{1}{N^2}(1-s_i)s_j^{(ij)}\,.
\label{simpleT}
\end{equation}
The form of this equation can be understood as follows: the $N^{-2}$ factor comes from selecting two sites (first $i$ then $j$) at random from the network; the factor of $(1-s_i)$ introduces some inertia on the part of agent $i$ -- the more satisfied they are with their present location, the less likely they are to move; the last factor of $s_j^{(ij)}$ is the attractiveness of site $j$ to the first agent selected -- the move is more likely if the destination site will provide greater satisfaction. \par
The combination of the network, initial condition, satisfaction function and transfer probabilities given here defines almost the bare bones of a Schelling-class model. The network has been abstracted away to a collection of disconnected pairs, the vacancies (introduced by Schelling simply as a device to get his agents moving \cite{Schelling1969}) have been removed,  the satisfaction function is reduced to picking a pair of numbers, and the transfer probabilities have an extremely simple form.
\subsection{Deterministic limit}
Write $x(t)$ for the interface density at time $t$, as defined in equation (\ref{def_x}). After analysing the possible changes to the system in one timestep, we will take a limit of large system size, in which a continuous time approximation is valid. As it is defined, $x(t)$ is a function of the state $\bmsigma$ of the whole system, however, the relationship between the two is sufficiently simple that it is possible to write expressions for the evolution of $x(t)$ that do not depend on $\bmsigma$.\par
\begin{figure}
\psfrag{a}{$\alpha$}
\psfrag{b}{$\beta$}
\centering\includegraphics[width=0.68\textwidth, trim=0 0 0 0]{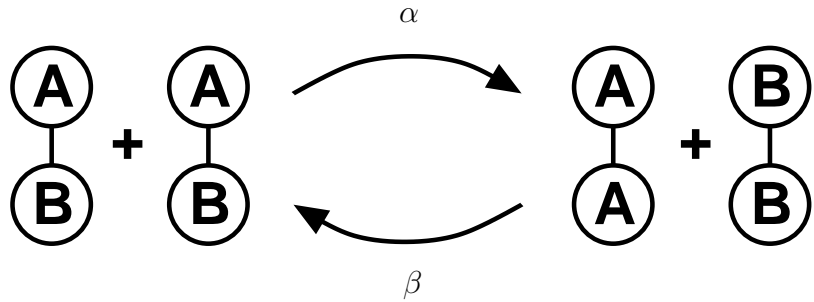}
\caption{A non-trivial change to the state of the simple Schelling variant discussed in the text occurs when a pair of heterogeneous edges become homogeneous, or vice versa. The rates $\alpha$ and $\beta$ are determined from the parameters of the model.}
\label{reaction}
\end{figure}

First note that, up to a trivial renaming of the sites, one system state of this simple model can differ from another only in regard of the number of agents which are paired with another of a different type. By enumerating the possible choices of agents to interact in one timestep, one finds that the only possibilities which lead to a change in the interface density are those in which two heterogeneous pairs swap agents to become homogeneous, or vice versa, as illustrated in Figure \ref{reaction}. These reactions result in a change of $\pm 4/N$ to the interface density and occur with probabilities given by (\ref{simpleT}).\par
The final step is to count the multiplicity of each possible arrangement. For instance, there are $Nx(t)/2$ heterogeneous pairs, and hence $Nx(t)\big(Nx(t)/2-1\big)$ ways of choosing two of these in order. Taking this result together with the above arguments and a similar calculation for homogeneous pairs, we find that the interface density evolves randomly with time according to 
\begin{equation}
\hspace{-2cm}x(t+1)=\cases{x(t)+\frac{4}{N}&with probability $\displaystyle\quad\Big(1-x(t)\Big)^2\,\frac{(1-u)v}{2}$\\ \\ x(t)-\frac{4}{N}&with probability $\displaystyle\quad x(t)\left(x(t)-\frac{2}{N}\right)\frac{(1-v)u}{2}$\\ \\ x(t) &otherwise.}
\label{TX}
\end{equation}
Thus, for this simple model at least, one does not require full knowledge of the system state in order to determine the probability law governing the interface density.\par
If one rescales time by a factor $N^{-1}$ and then takes the limit $N\to\infty$, analysis of equation (\ref{TX}) reveals that the interface density $x(t)$ approaches a deterministic continuous time function satisfying
\begin{equation}
\frac{dx}{dt}=\alpha \big(1-x(t)\big)^2 - \beta x(t)^2\,,\qquad x(0)=\frac{1}{2}\,,
\label{MF}
\end{equation}
where $\alpha=2(1-u)v$ and $\beta=2(1-v)u$. The initial condition $x(0)=1/2$ is simply the average interface density in the random initial condition we specified for the simple model. The values $\alpha$ and $\beta$ used above are transformed parameters determining the rate of creation and destruction of inhomogeneous edges.\par
The ordinary differential equation (\ref{MF}) is solvable, giving
\begin{equation}
x(t)=\frac{\sqrt{\big.\alpha\beta}+\alpha\tanh\Big(t\,\sqrt{\big.\alpha\beta}\Big)}{2\sqrt{\big.\alpha\beta}+\Big(\alpha+\beta\Big)\tanh\Big(t\,\sqrt{\big.\alpha\beta}\Big)}\,.
\label{MF_sol}
\end{equation}
This result is exact for the ensemble average in the large system limit. The equilibrium state is found by sending $t\to\infty$, giving
\begin{equation}
x(t)\,\longrightarrow\, \frac{\sqrt{\big.\alpha\beta}+\alpha}{\left(\sqrt{\alpha}+\sqrt{\beta}\,\right)^2} = \frac{1}{1+\sqrt{\beta/\alpha}}\,.
\label{equi}
\end{equation}
\subsection{Comparison with simulations}
As well as providing an exact result for the large system size limit of our highly simplified model, the equation (\ref{MF}) and its solution (\ref{MF_sol}) may be regarded as a simple `effective' theory for the emergence of segregation in more complicated Schelling-class models. The transformed parameters $\alpha$ and $\beta$ can be interpreted as the average rate of creation/destruction of inhomogeneous edges in the system. Considering the simplicity of the theory, manipulating these parameters to fit the curve (\ref{MF_sol}) to numerical data gathered from simulations is surprisingly successful. \par
\begin{figure}
\includegraphics[width=\textwidth, trim=0 0 0 0]{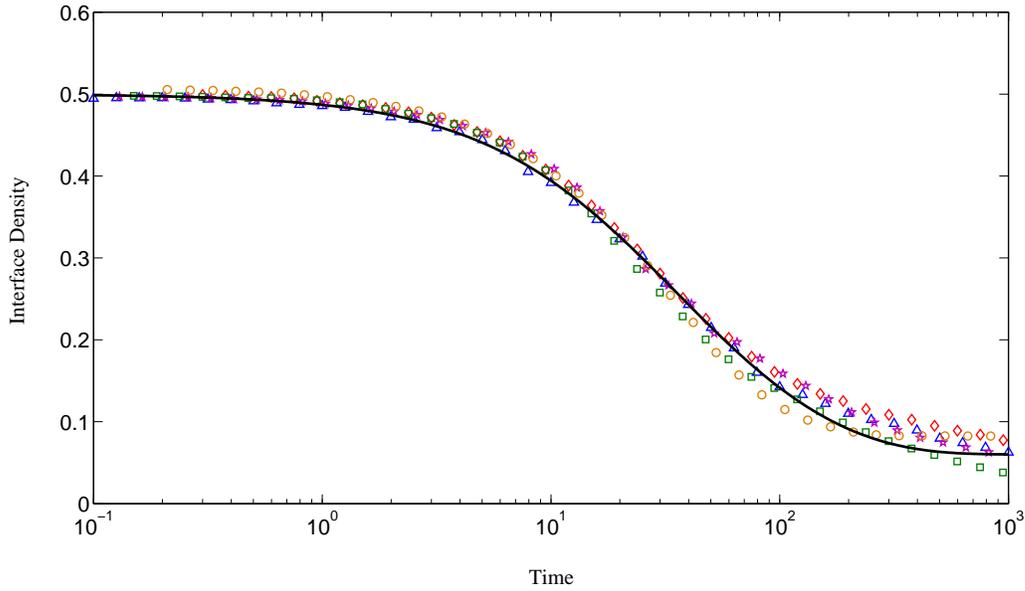}
\caption{Symbols -- time evolution of the interface density in a single simulation run of each of the example models discussed in the text, see Figure \ref{existing_models} for details. Solid line -- analytic result for interface density of the simplified model with parameters in equation (\ref{MF_sol}) chosen to approximately fit the simulation data.}
\label{fit}
\end{figure}
In Figure \ref{fit} we show the evolution of interface density in the five models considered in the previous section over four decades in time, overlaid with the analytic result for the simplified model, with parameters $\alpha=0.0002$ and $\beta=0.05$ chosen to approximately fit the simulation data. Translating these back to the model parameters $u$ and $v$, we find that the difference $u-v=0.0249$ is quite a bit smaller than the typical range of satisfaction values in the usual models. This can be understood by considering the effect of reducing the number of neighbours per site. In a typical model with many neighbours per site and starting from a random initial condition, most vacancies are surrounded by a mix of different agents and hence heterogeneous edges are created and destroyed at nearly the same rate, with only a slight imbalance due to agent preference. However this effect is not present in the simple model with only one neighbour per site, meaning that the timescale for segregation will be very much shorter, unless the parameters are adjusted to compensate.
\par
The curve (\ref{MF_sol}) can in fact be made to fit several of the models rather well for very much longer times, however, we have chosen not to show this. In the infinite network size limit $N\to\infty$, the long-time behaviour of the simple model is straightforward: if $\alpha>0$ the interface density approaches the non-zero equilibrium value (\ref{equi}), with the difference decaying as $e^{-2\sqrt{\alpha\beta}\,t}$; alternatively, if $\alpha=0$ then interface density scales as $x(t)\propto t^{-1}$. As discussed earlier, in simulations of the more complex models it is common to observe a scaling behaviour determined by the properties of the network and the dynamical rules, which will eventually be limited either by the process reaching a stable state, or by the effect of finite network size. The simplified model studied here was designed to investigate the initial emergence of segregation rather then the long-time dynamics of the different models, so a fit (no matter how good) to simulation data for much long times would not be very meaningful.\par
It is worth reiterating that the analysis presented here is intentionally limited; we have chosen one behavioural aspect of Schelling models and used a highly simplified model to help us study it. Whilst the interface density is only one of several interesting quantities which can be studied in Schelling-class models, it entirely characterises the simple model presented here, meaning that this model can be of little interest to those wishing to study other measures of segregation. Since there is no spatial dimension to the model, there are no interesting screenshots to exhibit, and no discussion to be had about pattern formation. Further, the extreme simplification made by providing each site with only one neighbour means that this model cannot be used to study the effects of interesting choices of satisfaction function (as in, for example \cite{Pancs2007}); the entire behaviour is determined by two parameters describing the rate of creation and destruction of heterogeneous edges. 
\par
We also point out that in many models one can, by exploring the extremes of the parameter space, find radically different behaviours which do not fit the characteristic pattern of Figure \ref{fit} and therefore cannot be reproduced by this simple model. This should not be seen as a drawback of the present approach however; the model was intentionally designed to investigate generic Schelling-class models in their `normal' range of behaviours, and not a specific type of model in an extreme setting. In fact, the approach advocated here, of tailoring the choice of model both to exhibit the phenomena one wishes to study and to be accessible to the theoretical tools available, can just as well be applied to these more unusual behaviours.
\section{Conclusion}
The interdisciplinary nature of research into Schelling's model has unfortunately led to a literature on the subject which lacks focus, with a great many different variations of the model having been proposed and simulated. Whilst simulation is certainly a powerful and important tool, a traditional theoretical physics approach based on the analytic solution of a judiciously chosen simplified model can still provide a significant contribution. In this article we have proposed a general scheme which we hope brings some order to the catalogue of model variations and can aid with the development of such theoretical analyses.\par
The principal goal of this work was to unite the many and varied generalisations to the model in a single, simple, mathematical framework. This was achieved by identifying four key components which together specify a Schelling-class model -- the network, the initial condition, the satisfaction function and the transfer probabilities. In each case these components satisfy certain heuristic rules to keep them in line with spirit of Schelling's work, whilst being general enough to include almost all variants currently existing in the literature. \par
We also discussed some of the various statistics which can be used to chart the emergence of segregation in these models, choosing here to focus on the interface density as a representative measure. It is worth commenting that although quantities such as interface density can provide useful insights into the behaviour of a computation model, a great deal of care must be taken if one hopes to infer something about segregation in real societies. To quote Schelling, \textit{quantitative measures, of course, refer exclusively to an artificial checkerboard and are unlikely to have any quantitative analogue in the living world} \cite{Schelling1969}. See also \cite{Edmonds2005} for a related discussion on the interpretation of the model. Moreover, the problem of quantifying segregation in real populations is itself an open question amongst social scientists. It is beyond the scope of this article to comment on this debate, we instead refer the interested reader to \cite{White1986,Simpson2007}.\par
One immediate benefit of approaching the plethora of variations to the model in the systematic way developed in this article is that it allows for a structured discussion of the changes made from one iteration to the next. For example, we see that moving from randomly jumping agents which are popular in the physics literature \cite{DallAsta2008,Gauvin2009,Gauvin2010} to best-response dynamics favoured by economists \cite{Fagiolo2007,Pancs2007} does not result in a qualitative change to the way in which segregation develops. Simulation results for the time evolution of interface density in five different models taken from several disciplines were presented, each displaying the same characteristic form in short to medium timescales. \par
The second useful application of the unified framework proposed here is in developing analytically tractable models. When a particular phenomenon is observed in several different models, careful consideration of the role of the different components used in these models can point the way to simpler versions which will be amenable to existing theoretical tools. To demonstrate this methodology we presented the creation and analysis of a highly abstract form of the model, designed specifically to provide a solvable prediction for the characteristic time dependence of interface density which was observed in the earlier simulations. The analysis resulted in a very simple effective theory for interface density for Schelling-class models, depending on two parameters describing the rate of creation and destruction of homogeneous edges. These parameters can be chosen to provide to fit simulation data from the short to medium timescales of the more complex models.\par
Whilst the mathematical framework we have developed is designed to be as comprehensive as possible, the same is not true of the analytical work presented here, which is only a demonstration of what is possible. There are many interesting phenomena which have manifested in simulations of Schelling-class models, and we anticipate that most of these will be accessible to analytic treatment of carefully designed models. Examples of future research directions include: jamming transitions when too few vacancies are present to allow agents to move as desired; pattern formation in which the agents spontaneously form bands of alternating types; coarsening processes and domain growth; solid/liquid transitions depending on the likelihood of movement of satisfied agents; the influence of stochasticity and finite size effects. In a forthcoming paper we will show how patch-based models provide novel and analytically tractable examples of jamming and pattern formation \cite{Rogers2011iii}. \par

\section*{Acknowledgements} 
This work was funded by the EPSRC under grant number EP/H02171X/1.  

\bibliographystyle{iopart-num}
\bibliography{Schelling}
\end{document}